\documentclass[aps,prl,superscriptaddress,twocolumn,floatfix,showpacs,citeautoscript,longbibliography,hyperlinks]{revtex4-1}

\usepackage[english]{babel}
\usepackage{amsmath,amsfonts,bm,graphicx,verbatim,mathrsfs,units,color}
\usepackage[colorlinks=true,citecolor=blue]{hyperref}

\newcommand{\revision}[1]{{{#1}}}

\definecolor{ThesisBlue}{RGB}{40,140,170}
\definecolor{Grey}{RGB}{100,100,100}

\begin{document}

\title{Electron Waiting Times of a Cooper Pair Splitter}
\author{Nicklas Walldorf}
\affiliation{Center for Nanostructured Graphene (CNG), Department of Micro- and Nanotechnology, Technical University of Denmark, DK-2800 Kongens Lyngby, Denmark}
\author{Ciprian Padurariu}
\affiliation{Department of Applied Physics, Aalto University, 00076 Aalto, Finland}
\author{Antti-Pekka Jauho}
\affiliation{Center for Nanostructured Graphene (CNG), Department of Micro- and Nanotechnology, Technical University of Denmark, DK-2800 Kongens Lyngby, Denmark}
\author{Christian Flindt}
\affiliation{Department of Applied Physics, Aalto University, 00076 Aalto, Finland}
\date{\today}

\begin{abstract}
Electron waiting times are an important concept in the analysis of quantum transport in nano-scale conductors. Here we show that the statistics of electron waiting times can be used to characterize Cooper pair splitters that create spatially separated spin-entangled electrons.  A short waiting time between electrons tunneling into different leads is associated with the fast emission of a split Cooper pair, while long waiting times are governed by the slow injection of Cooper pairs from a superconductor. Experimentally, the waiting time distributions can be measured using real-time single-electron detectors in the regime of slow tunneling, where conventional current measurements are demanding. Our work is important for understanding the fundamental transport processes in Cooper pair splitters and the predictions may be verified using current technology.
\end{abstract}

\maketitle

\textit{Introduction.---}
\revision{Quantum technologies that exploit non-classical phenomena such as the discreteness of physical observables, coherent superpositions, and quantum entanglement} promise solutions to current challenges in communication, computation, sensing, and metrology~\cite{zagoskin:2011}. For solid-state quantum computers, an important building block is a device that can generate pairs of entangled electrons~\cite{Ladd:2010}. In one prominent approach, Cooper pairs in a superconductor are converted into spatially separated electrons that preserve the entanglement of their spins~\cite{Lesovik2001,Recher2001}. Cooper pair splitters have been realized in architectures based on superconductor--normal-state hybrid systems \cite{Beckmann2004,Russo2005,Deacon:Cooper}, InAs nanowires \cite{Hofstetter:Cooper,Hofstetter2011,Das2012,Fulop2015}, carbon nanotubes \cite{Herrmann:Carbon,Herrmann:Spectroscopy,Schindele:Near,Fueloep:Local,Schindele:Nonlocal}, and recently graphene structures \cite{Tan:Cooper,Borzenets:High,Islam2017}.

\begin{figure}
  \centering
  \includegraphics[width=1\columnwidth]{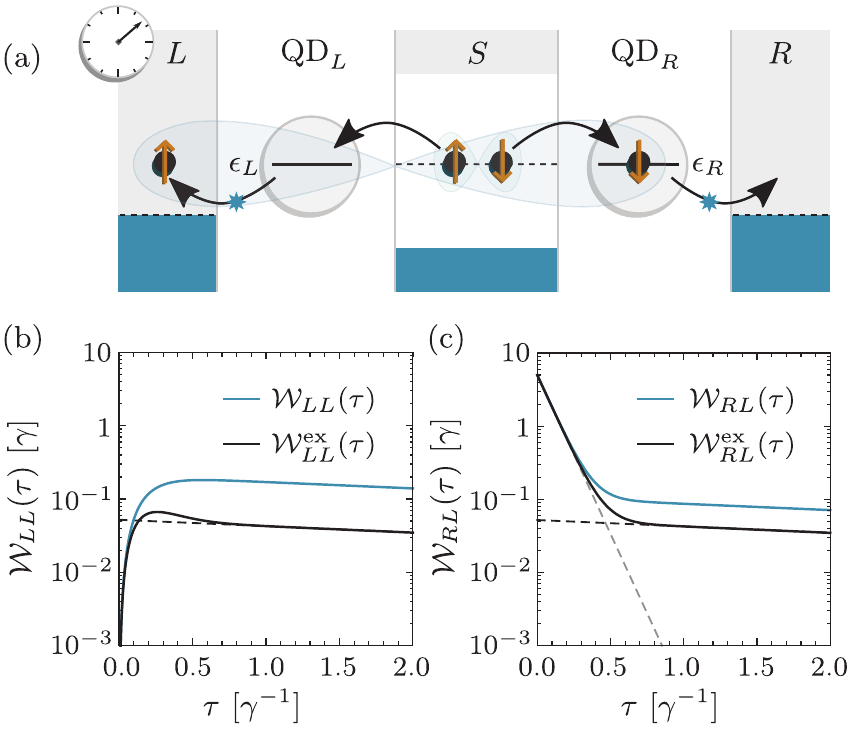}
  \caption{Electron waiting times of a Cooper pair splitter. (a) Two QDs are coupled to a superconducting source of Cooper pairs and two normal-metal drains. A tunneling event (star) starts the clock\revision{, which symbolizes the measurement scheme based on single-electron detectors \cite{Gustavsson2009,Maisi:Real,Maisi:Full,Gorman2017}}. A subsequent tunneling event stops it. WTDs for tunneling into the same/different leads are shown in panels (b) and (c). The WTDs $\mathcal{W}_{ji}(\tau)$ $[\mathcal{W}^{\mathrm{ex}}_{ji}(\tau)]$ are evaluated using Eq.~(\ref{eq:wtd}) [(\ref{eq:ex_wtd})]. Parameters are $\xi:=\gamma_L=\gamma_R=10\gamma$, $\gamma_\text{CPS}=\gamma_\text{EC}=\gamma$, and $\epsilon_L=\epsilon_R=0$. Dashed lines are exponentials with decay rates $\xi$ (grey) and $2 \gamma_\text{CPS}^2/\xi$~(black). \revision{Corresponding to the recent experiments, the rate $\gamma$ would be on the order of kilo-hertz, and the waiting times would be in the millisecond range \cite{Maisi:Real,Maisi:Full,Gorman2017}.}}
\label{fig:Fig1}
\end{figure}

The efficiency of Cooper pair splitters can be determined using conductance measurements \cite{Hofstetter:Cooper,Hofstetter2011,Das2012,Fulop2015,Herrmann:Carbon,Herrmann:Spectroscopy,Schindele:Near,Fueloep:Local,Schindele:Nonlocal,Tan:Cooper}. For some setups, the efficiency is approaching unity~\cite{Schindele:Near,Das2012}, indicating that Cooper pair splitters may be suited for electronics-based quantum technologies. One may now hope to detect the entanglement of the outgoing electrons by measuring the cross-correlations of the currents in the output channels \cite{Kawabata2001,Chtchelkatchev2002,Sauret:Spin,Das2012}. However, while these approaches are based on conventional current measurements, recent progress in the real-time detection of single electrons is opening another promising avenue for understanding quantum transport in nano-scale devices \cite{Gustavsson2009}.

In this Letter we propose to characterize Cooper pair splitters using the distribution of electron waiting times. The electron waiting time is the time that passes between subsequent tunneling events. Waiting time distributions (WTDs) have in recent years been investigated theoretically for quantum transport in quantum dots~\cite{Brandes2008,Welack2008,Welack2009,Albert2011,Thomas2013,Tang2014,Tang2014Full,Sothmann2014,SeoaneSouto2015,Talbo2015,Rudge2016,Rudge20162,Ptaszynski2017,Potanina2017,Kosov2017}, mesoscopic conductors~\cite{Albert2012,Haack2014,Thomas2014,Dasenbrook2014,Albert2014,Dasenbrook:Electron,Dasenbrook2016,Hofer2016}, and superconducting devices~\cite{Rajabi2013,Dambach2015,Dambach2016,Albert2016,Chevallier2016}. \revision{Moreover, in a very recent experiment, the distribution of electron waiting times was measured for a quantum dot~\cite{Gorman2017}.} Here, we show that the WTD is a sensitive tool to understand the working principle of the Cooper pair splitter in Fig.~\ref{fig:Fig1}(a). As we discuss below, WTDs such as those in Fig.~\ref{fig:Fig1}(b) and (c) provide clear signatures of the Cooper pair splitting. Specifically, the splitting of Cooper pairs is associated with a large peak at short times in the WTD for tunneling into different drains, Fig.~\ref{fig:Fig1}(c). \revision{This information is complementary to what can be learned from conventional current and noise measurements. In addition}, with the ability to detect single electrons participating in Andreev tunneling across normal-state--superconductor interfaces \cite{Maisi:Real,Maisi:Full}, a measurement of the electron waiting times in a Cooper pair splitter appears feasible with current technology. \revision{More precisely, in the recent experiment on WTDs, the typical waiting times were on the order of milliseconds~\cite{Gorman2017}, which corresponds well to the kilo-hertz tunneling rates reported in Refs.~\cite{Maisi:Real,Maisi:Full}. Importantly, such low tunneling rates do not produce electrical currents that can be measured using standard techniques. On the other hand, the tunneling of electrons can be detected in real-time, and the distribution of waiting times can be measured.}

\textit{Cooper pair splitter.---} The Cooper pair splitter consists of two quantum dots (QDs) coupled to a superconductor and two normal leads~\cite{Recher2001}. The grounded superconductor acts as a source of Cooper pairs. The negatively biased leads serve as drains for electrons in the QDs. Coulomb interactions are so strong that each QD cannot be occupied by more than one electron at a time. With a large superconducting gap, we may focus on the subgap transport (the working regime is specified below). The superconductor can then be included in an effective Hamiltonian of the QDs reading~\cite{Sauret:Quantum,Eldridge:Superconducting,Braggio2011,Trocha2015,Amitai2016,Hussein2016,Hussein2017}
\begin{equation}
\begin{split}
\hat{H}_{\mathrm{QDs}}=&\sum_{\ell\sigma}\epsilon_\ell^{\phantom\dagger}\hat{d}_{\ell\sigma}^\dagger \hat{d}_{\ell\sigma}^{\phantom\dagger}-\gamma_\text{EC}\sum_\sigma\!\left(\hat{d}_{L\sigma}^\dagger \hat{d}_{R\sigma}^{\phantom\dagger}+\text{h.c.}\!\right)\!\\
&-\frac{\gamma_\text{CPS}}{\sqrt{2}}\!\left(\hat{d}_{L\downarrow}^\dagger \hat{d}_{R\uparrow}^\dagger-\hat{d}_{L\uparrow}^\dagger \hat{d}_{R\downarrow}^\dagger\!+\text{h.c.}\right),
\end{split}
\end{equation}
Here, the operator $\hat{d}_{\ell\sigma}^\dagger$ ($\hat{d}_{\ell\sigma}^{\phantom\dagger}$) creates (annihilates) an electron in $\text{QD}_\ell$, $\ell\in\{L,R\}$ with spin $\sigma\in\{\uparrow,\downarrow\}$ and energy $\epsilon_\ell$ relative to the chemical potential of the superconductor, $\mu_S=0$. The amplitudes $\gamma_\text{CPS}$ and $\gamma_\text{EC}$ correspond to Cooper pair splitting (CPS) and elastic cotunneling (EC) processes, respectively\revision{, and can be expressed in terms of microscopic parameters following Ref.~\cite{Sauret:Quantum}. We have excluded direct coupling between the QDs as in the experiment of Ref.~\cite{Tan:Cooper}, but such processes can easily be incorporated within our formalism}. In the CPS processes, a Cooper pair in the superconductor is converted into two spin-entangled electrons in a singlet state with one electron in each QD, or vice versa. Such processes are favored when the empty state of the QDs is energetically degenerate with the doubly occupied state, $\epsilon_L+\epsilon_R=0$ \cite{Chevallier:Current,Hiltscher2011,Floeser2013,Sadovskyy2015}. In the spin-preserving EC processes, an electron in one of the QDs is transferred via the superconductor to the other QD. These processes are on resonance when the QD levels are energetically aligned, $\epsilon_L=\epsilon_R$.

Transport through each QD is described by resonant tunneling and must be treated to all orders in the coupling to the leads. When the resonant level is deep inside the transport energy window, the transport can be described by a Markovian quantum master equation for the reduced density matrix $\hat{\rho}$ of the QDs (with $\hbar=1$) \cite{Sauret:Quantum,Gurvitz2016}
\begin{equation}
\frac{d}{dt}\hat{\rho}=\mathcal{L}\hat{\rho}=-i[\hat{H}_{\mathrm{QDs}},\hat{\rho}]+\mathcal{D}\hat{\rho}.
\label{eq:me}
\end{equation}
Here, the Liouvillian $\mathcal{L}$ describes both coherent processes governed by $\hat{H}_{\mathrm{QDs}}$, and incoherent single-electron jumps to the normal metals captured by the Lindblad dissipator
\begin{equation}
\mathcal{D}\hat{\rho}=\sum_{\ell\sigma}\gamma_\ell^{\phantom\dagger}\!\left[\hat{d}_{\ell\sigma}^{\phantom\dagger}\hat{\rho} \hat{d}_{\ell\sigma}^\dagger-\frac{1}{2}\{\hat{\rho},\hat{d}_{\ell\sigma}^\dagger \hat{d}_{\ell\sigma}^{\phantom\dagger}\}\right].
\label{eq:diss}
\end{equation}
We take the rate $\gamma_\ell$  at which electrons leave via lead $\ell$ to be independent of the spin. To summarize, we work in the regime $U, \Delta \gg |V|\gg \epsilon_\ell, \gamma_\ell,\gamma_\text{CPS},\gamma_\text{EC}$, where $U$ is the Coulomb interaction energy, $\Delta$ is the superconducting gap, and $V$ is the negative voltage. Due to the large negative bias, the electron transport from the QDs to the drain electrodes is unidirectional and the thermal smearing of the distribution functions in the leads becomes unimportant. 
\textit{Electron waiting times.---}
We characterize the Cooper pair splitter by the distribution of electron waiting times. Given that an electron with spin $\sigma$ has just tunneled into lead $\ell$, the electron waiting time $\tau$ is the time that passes until another electron with spin $\sigma'$ tunnels into lead $\ell'$. The electron waiting time is a fluctuating quantity that must be characterized by a probability distribution. The terms in Eq.~(\ref{eq:diss}) of the form $\mathcal{J}_{\ell\sigma}\hat{\rho}\equiv\gamma_\ell^{\phantom\dagger}\hat{d}_{\ell\sigma}^{\phantom\dagger}\hat{\rho} \hat{d}_{\ell\sigma}^\dagger$ describe incoherent tunneling processes in which an electron with spin $\sigma$ in $\text{QD}_\ell$ tunnels into lead $\ell$. The distribution of waiting times between transitions of type $i=\ell\sigma$ and $j=\ell'\sigma'$  can then be expressed  as~\cite{Brandes2008,Dasenbrook:Electron,Carmichael:Photoelectron}
\begin{equation}
\mathcal{W}_{ji}(\tau)=\frac{\text{Tr}[\mathcal{J}_j e^{(\mathcal{L}-\mathcal{J}_j)\tau}\mathcal{J}_{i}\hat{\rho}_S]}{\text{Tr}[\mathcal{J}_{i}\hat{\rho}_S]},
\label{eq:wtd}
\end{equation}
where $\hat{\rho}_S$ is the stationary density matrix given as the normalized solution to the equation $\mathcal{L}\hat{\rho}_S=0$. The expression above for the WTD can be understood as follows: after a transition of type $i$ has occurred, the system is evolved until the next transition of type $j$ happens. The denominator ensures that the WTD is normalized to unity when integrated over all possible waiting times.

Figures~\ref{fig:Fig1}(b) and~\ref{fig:Fig1}(c) show WTDs for transitions into the same lead and different leads, respectively. Experimentally, transitions between different charge states can be monitored using charge detectors that measure the occupation of each QD \cite{Gustavsson2009,Gorman2017,Maisi:Real,Maisi:Full}. In Fig.~\ref{fig:Fig1}(b), we consider the waiting time between transitions into the left lead. Here, the coupling to the drain electrodes is much larger than the coupling to the superconductor, $\gamma_L,\gamma_R\gg\gamma_\text{CPS},\gamma_\text{EC}$. As the QDs cannot be doubly-occupied, the WTD is strongly suppressed at short times, $\tau\ll\gamma_\text{CPS}^{-1}$, and vanishes completely at $\tau=0$, since simultaneous transitions into the same lead are not possible. At long times, the WTD is governed by the slow refilling of the left QD  and the subsequent tunneling of an electron into the left lead. This WTD resembles what one would expect for single-electron tunneling through a single QD without any Cooper pair splitting \cite{Brandes2008}.

A very different picture emerges from the waiting time between transitions into different leads. In Fig.~\ref{fig:Fig1}(c), the splitting of a Cooper pair is signaled by a large peak in the WTD at short times, $\tau\ll\gamma_\text{CPS}^{-1}$. In this case, the tunneling of an electron into the left lead is quickly followed by a transition into the right lead on a time-scale given by the coupling to the right lead,~$\gamma_R^{-1}$. The slow decay of the WTD describes the waiting time between electrons originating from different Cooper pairs. This WTD clearly reflects the non-local nature of the CPS processes and it carries information about the \emph{short} waiting times between electrons from the \emph{same} Cooper pair and the \emph{long} waiting times between electrons originating from \emph{different} Cooper pairs. Experimentally, a measurement of the WTD in Fig.~\ref{fig:Fig1}(c) would constitute a strong evidence of efficient Cooper pair splitting.

\begin{figure}
  \centering
  \includegraphics[width=0.95\columnwidth]{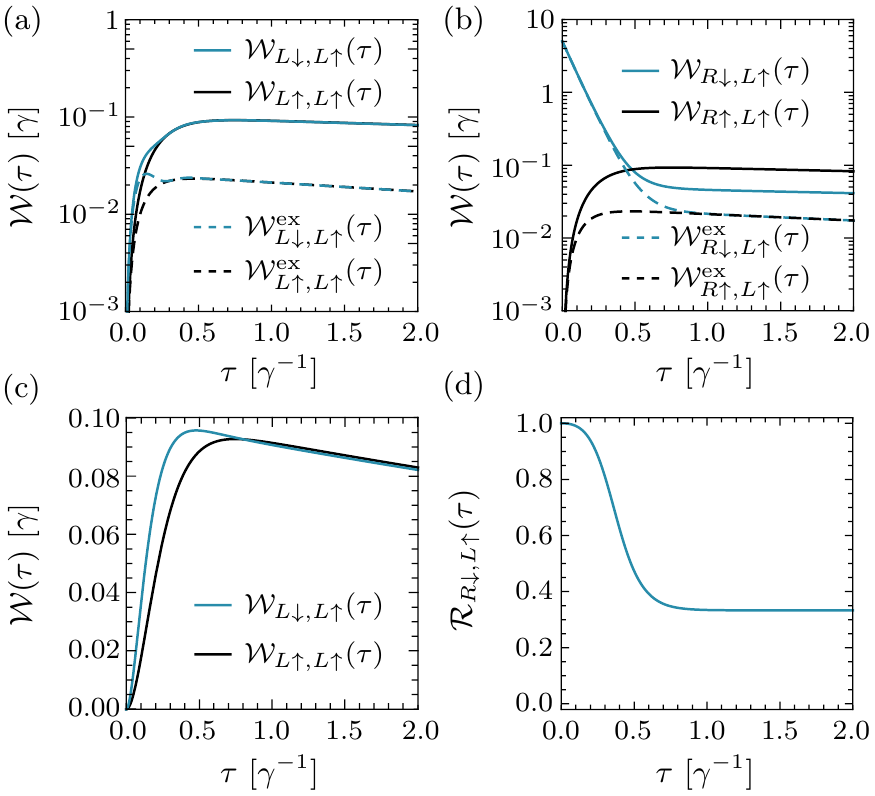}
  \caption{Spin-resolved WTDs. (a) Spin-revolved WTDs for tunneling into the same lead. (b) Spin-revolved WTDs for tunneling into different leads. In (a) and (b), the parameters are $\gamma_L=\gamma_R\equiv10\gamma$, $\gamma_\text{CPS}=\gamma_\text{EC}=\gamma$, and $\epsilon_L=-\epsilon_R=10\gamma$. (c) Spin-revolved WTDs for tunneling into the same lead with same parameters except that $\epsilon_L=\epsilon_R=0$. (d) The branching ratio in Eq.~(\ref{eq:branch_ratio}) corresponding to the WTDs in panel (b).}
\label{fig:Fig2}
\end{figure}

\revision{\textit{Exclusive WTDs.---}} To better understand the time-scales that enter the WTDs, we introduce \emph{exclusive} WTDs. Again, we consider the waiting time that passes between transitions of  types $i$ and $j$. However, we now exclude cases, where any other transitions occur during the waiting time. This WTD is then defined as~\cite{Brandes2008,Rajabi2013}
\begin{equation}
\mathcal{W}^{\mathrm{ex}}_{ji}(\tau)=\frac{\text{Tr}[\mathcal{J}_j e^{\mathcal{L}^{\mathrm{ex}}\tau}\mathcal{J}_{i}\hat{\rho}_S]}{\text{Tr}[\mathcal{J}_{i}\hat{\rho}_S]},
\label{eq:ex_wtd}
\end{equation}
where $\mathcal{L}^{\mathrm{ex}}=\mathcal{L}-\sum_k \mathcal{J}_k$ removes all possible transitions from the full time evolution given by $\mathcal{L}$. In contrast to the WTD in Eq.~(\ref{eq:wtd}), the exclusive WTD is only normalised upon integrating over all waiting times \textit{and} summing over all types of final events. Due to its simpler structure, the exclusive WTD can be evaluated analytically. For example, with $\gamma_L=\gamma_R=\xi$ and $\epsilon_L=-\epsilon_R=\epsilon$, we find
\begin{equation}\label{eq:cexwt}
\begin{split}
\mathcal{W}_{\ell\sigma,\ell'\sigma}^\text{ex}(\tau)=&\frac{\xi }{2}e^{-\xi\tau}\alpha_\text{CPS}^2\!\left[1-\cos\left(\omega_\text{CPS}\tau\right)\right],\\
\mathcal{W}_{\ell\sigma,\ell\bar{\sigma}}^\text{ex}(\tau)=&\xi e^{-\xi\tau}\alpha_\text{EC}^2\!\left[1-\cos(\omega_\text   {EC}\tau)\right]+\mathcal{W}_{\ell\sigma,\ell\sigma}^\text{ex}(\tau),\\
\mathcal{W}_{\ell\sigma,\bar{\ell}\bar{\sigma}}^\text{ex}(\tau)=&\frac{\xi }{2}e^{-\xi\tau}+2\mathcal{W}_{\ell\sigma,\ell\sigma}^\text{ex}(\tau)-\mathcal{W}_{\ell\sigma,\ell\bar{\sigma}}^\text{ex}(\tau),
\end{split}
\end{equation}
with $\bar{L}=R$ and $\bar{\uparrow}=\downarrow$ and vice versa, and we have identified the frequencies $\omega_\text{CPS} = 2\sqrt{\gamma_\text{CPS}^2-(\xi/2)^2}$ and $\omega_\text{EC}= 2\sqrt{\gamma_\text{EC}^2+\epsilon^2}$ associated with the coherent CPS and EC processes and introduced the ratios $\alpha_\text{CPS}=\gamma_\text{CPS}/\omega_\text{CPS}$ and $\alpha_\text{EC}=\gamma_\text{EC}/\omega_\text{EC}$. If $\gamma_\text{CPS}\gg \gamma_L,\gamma_R$, the WTD exhibits oscillations with frequency $\omega_\text{CPS}\simeq 2\gamma_\text{CPS}$. By contrast, for $\gamma_\text{CPS}\ll \gamma_L,\gamma_R$, the frequency becomes imaginary and now rather corresponds to an exponential decay. In Fig.~\ref{fig:Fig1}, we show the exclusive WTDs $\mathcal{W}_{\ell\ell'}^\text{ex}(\tau)=\sum_{\sigma,\sigma'}\mathcal{W}_{\ell\sigma,\ell'\sigma'}^\text{ex}(\tau)/2$. \revision{For short times, we have $\mathcal{W}_{LL}^\text{ex}(\tau)\simeq (\omega_\text{CPS}\tau)^2$. By contrast, for the WTD in Fig.~\ref{fig:Fig1}(c) the short-time behavior $\mathcal{W}_{RL}^\text{ex}(\tau) \simeq e^{-\xi\tau}$ is governed by the escape rate, while the long-time decay $\mathcal{W}_{RL}^\text{ex}(\tau) \simeq e^{-2 \tau\gamma_\text{CPS}^2/\xi}$ also involves the CPS amplitude.}

\textit{Spin-resolved WTDs.---} The splitting of Cooper pairs can be identified in the charge-resolved WTDs as we saw in Fig.~\ref{fig:Fig1}(c). Still, further information can be obtained from the spin-resolved WTDs. Experimentally, one might measure spin-resolved WTDs using ferromagnetic detectors \cite{Trocha2015,Malkoc:Full,Busz2017,Wrzesniewski2017}. In Fig.~\ref{fig:Fig2}, we show WTDs that are resolved with respect to the spin degree of freedom. In Figs.~\ref{fig:Fig2}(a) and (b), the levels are detuned so that only CPS processes are on resonance. Again, the WTDs for transitions into the same lead show essentially no signatures of the CPS processes. By contrast, the CPS processes can be identified in the WTD in Fig.~\ref{fig:Fig2}(b) for transitions into different leads. Here, the CPS processes show up as a large enhancement at short times in the WTD for opposite spins. Due to the splitting of a Cooper pair, the tunneling of a spin-up electron into the left lead is likely followed by the tunneling of a spin-down electron into the right lead. A similar enhancement is not found for electrons with the same spin, since they must originate from different Cooper pairs.

\begin{figure}
  \centering
  \includegraphics[width=0.95\columnwidth]{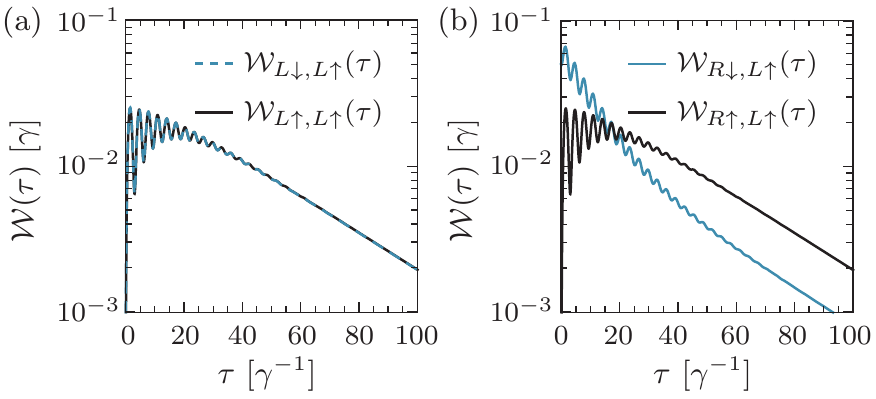}
  \caption{Coherent oscillations. (a) Spin-revolved WTDs for tunneling into the same lead. (b) Spin-revolved WTDs for tunneling into different leads. In both panels, the parameters are $\gamma_L=\gamma_R\equiv0.1\gamma$, $\gamma_\text{CPS}=\gamma_\text{EC}=\gamma$, and $\epsilon_L=-\epsilon_R=10\gamma$.}
\label{fig:Fig3}
\end{figure}

In Fig.~\ref{fig:Fig2}(c), both the CPS and EC processes are tuned into resonance. The combination of these processes lead to an enhancement at intermediate times in the WTD for electrons with opposite spins tunneling into the same lead. In this case, two electrons from a Cooper pair can exit into the same drain due to a spin-preserving EC process that transfers the second electron from the right to the left QD before it exits via the left drain. This is not possible for electrons with the same spin, since they cannot originate from the same Cooper pair, and the corresponding WTD is not enhanced in a similar way.

Importantly, from the spin-resolved WTDs, we can evaluate the branching ratio of the spins defined as
\begin{equation}
\mathcal{R}_{R\downarrow,L\uparrow}(\tau)\equiv\frac{\mathcal{W}_{R\downarrow,L\uparrow}(\tau)}{\mathcal{W}_{R\downarrow,L\uparrow}(\tau)+\mathcal{W}_{R\uparrow,L\uparrow}(\tau)}.
\label{eq:branch_ratio}
\end{equation}
The branching ratio is the probability that two electrons, which tunnel into different leads separated by the waiting time $\tau$, have opposite spins. Figure \ref{fig:Fig2}(d) shows that it is highly probable that electrons separated by a short waiting time have oppositive spins and they likely originate from the same Cooper pair. This finding is important since it allows us to conclude that the large peak in Fig.~\ref{fig:Fig1}(c) with near-unity probability corresponds to opposite spins originating from the same Cooper pair~\cite{Scheruebl2014}.

Until now, we have assumed that the coupling to the drains is much larger than the coupling to the superconductor. This regime may be most attractive for efficient Cooper pair splitting, since the split pair of electrons is quickly transferred to the drains. However, the opposite regime, $\gamma_\text{CPS},\gamma_\text{EC}\gg\gamma_L,\gamma_R$, is also interesting. In Fig.~\ref{fig:Fig3}, \revision{the rate of escape to the drains is so slow that several coherent oscillations between the QDs and the superconductor can be completed \cite{Brandes2008,Thomas2013,Rajabi2013}. As discussed after Eq.~(\ref{eq:cexwt}), the frequency of the oscillations is given by $\omega_\text{CPS}\simeq 2\gamma_\text{CPS}$.}

\textit{Joint WTDs.---} The WTDs concern waiting times between subsequent tunneling events. However, they do not describe correlations between consecutive waiting times. Such correlations can be characterized by evaluating the joint distribution of electron waiting times \cite{Dasenbrook:Electron,Dambach2016,Zhang2017}
\begin{equation}\label{eq:jcwt}
\mathcal{W}_{kji}(\tau_1,\tau_2)=\frac{\text{Tr}[\mathcal{J}_{k}e^{(\mathcal{L}-\mathcal{J}_k)\tau_2}\mathcal{J}_j e^{(\mathcal{L}-\mathcal{J}_j)\tau_1}\mathcal{J}_{i}\hat{\rho}_S]}{\text{Tr}[\mathcal{J}_{i}\hat{\rho}_S]},
\end{equation}
which generalizes Eq.~(\ref{eq:wtd}) to subsequent waiting times between transitions of type $i$, $j$, and $k$. For uncorrelated waiting times, the joint distribution factorizes as $\mathcal{W}_{kj}(\tau_2)\mathcal{W}_{ji}(\tau_1)$ in terms of the individual WTDs. Correlations can be quantified by the correlation function
\begin{equation}
\Delta\mathcal{W}_{kji}(\tau_1,\tau_2)=\frac{\mathcal{W}_{kji}(\tau_1,\tau_2)-\mathcal{W}_{kj}(\tau_2)\mathcal{W}_{ji}(\tau_1)}{\mathcal{W}_{kj}(\tau_2)\mathcal{W}_{ji}(\tau_1)},
\end{equation}
which is positive (negative) for positively (negatively) correlated waiting times and zero without correlations.

\begin{figure}[!t]
  \centering
  \includegraphics[width=0.95\columnwidth]{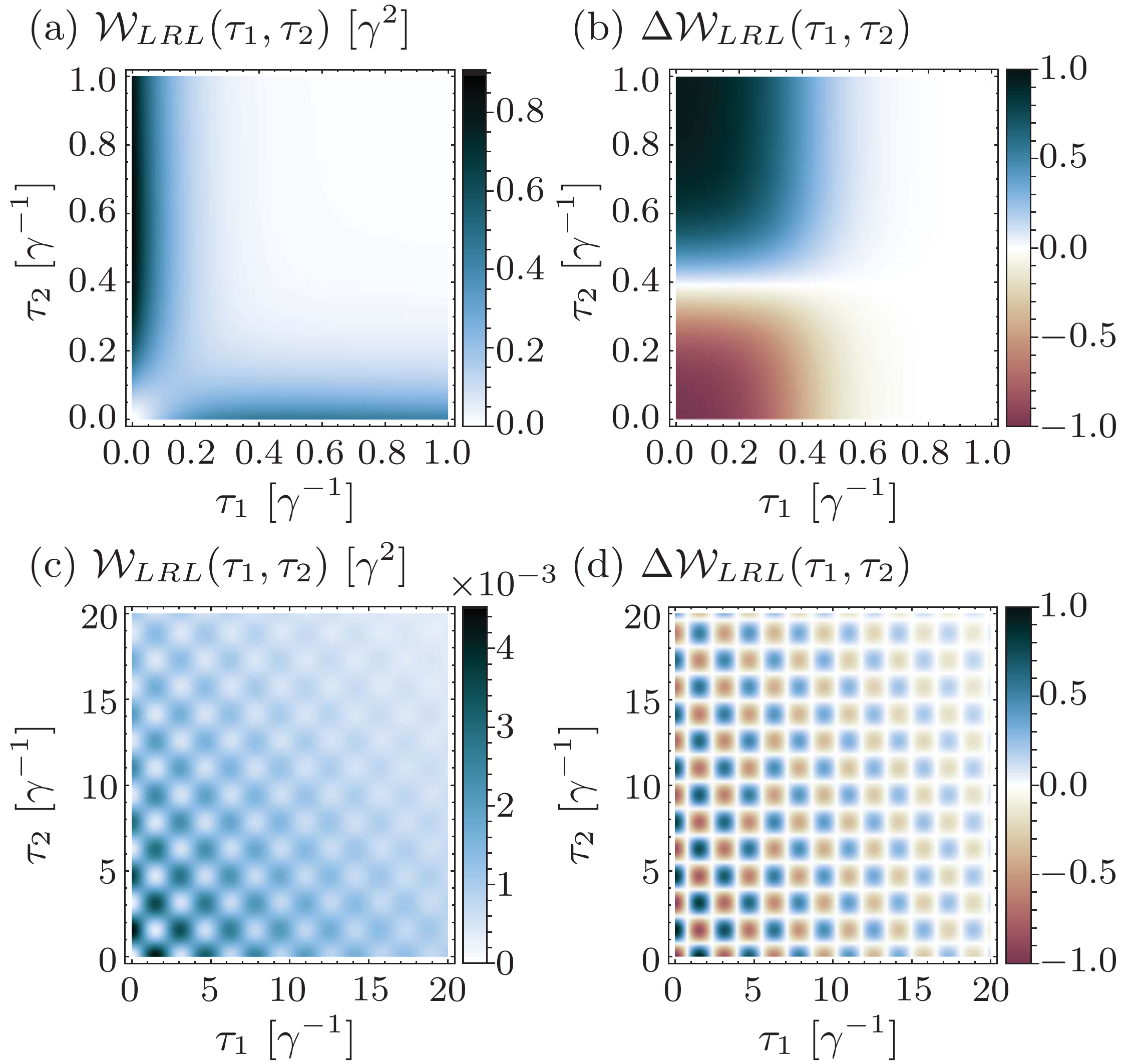}
  \caption{Joint WTDs and correlation functions. In panels (a) and (b), the parameters are $\gamma_L=\gamma_R=10\gamma$, $\gamma_\text{CPS}=\gamma_\text{EC}\equiv\gamma$, and $\epsilon_L=\epsilon_R=0$. In panels (c) and (d), the parameters are $\gamma_L=\gamma_R=0.1\gamma$, $\gamma_\text{CPS}=\gamma_\text{EC}=\gamma$, and $\epsilon_L=\epsilon_R=0$.}
\label{fig:Fig4}
\end{figure}

Figure~\ref{fig:Fig4} shows joint WTDs \revision{and correlation functions} for electrons arriving in different leads. In panels~(a)~and~(b), the coupling to the drains is much larger than the coupling to the superconductor. \revision{We see that a short waiting time is likely followed by a long waiting time, but unlikely followed by another short waiting time. A short waiting time corresponds to two electrons originating from the same Cooper pair, while a long waiting time is given by the slow refilling of the QDs by a split Cooper pair. The observed correlations reflect that the two processes, i.e.~emission into the drains and refilling from the superconductor, follow one after another. A similar behavior is seen in panels (c) and (d), where the coupling to the superconductor is the largest. However, now the rate of escape to the drains is so slow that coherent oscillations between the QDs and the superconductor have time to form, giving rise to the oscillatory pattern in the joint WTD and the correlation function.}

\textit{Conclusions.---} We have proposed to use waiting time distributions to characterize Cooper pair splitters.  The non-local nature of the Cooper pair splitting can be clearly identified in the distribution of waiting times. Based on the recent progress in the real-time detection of Andreev tunneling, we expect the predictions to be accessible in future experiments. Specifically, a measurement of the WTD would constitute a strong evidence of efficient Cooper pair splitting in the regime of slow tunneling, where conventional current measurements are demanding. Theoretically, it would be interesting to formulate a Bell-like inequality for the waiting times to certify the entanglement of the split Cooper pairs.

\acknowledgements
\emph{Acknowledgements.---} We thank P.~J.~Hakonen, G.~B.~Lesovik, M.~V.~Moskalets, J.~P.~Pekola, and B.~Sothmann for valuable discussions. Authors at Aalto are associated with Centre for Quantum Engineering. The Center for Nanostructured Graphene is sponsored by the Danish Research Foundation (Project DNRF103).

%\bibliography{journalabbreviations,references}
%

\end{document}